# Joint Optimization of Scheduling and Routing in Multicast Wireless Ad-Hoc Network Using Soft Graph Coloring and Non-linear Cubic Games

Ebrahim Karami[1], *Member, IEEE* and Savo Glisic[2], *Senior Member, IEEE*

[1]Department of Systems and Computer Engineering, Carleton University, Ottawa, ON, Canada
[2]Centre for Wireless Communications (CWC), University of Oulu, Oulu, Finland

*Abstract—* In this paper we present matrix game-theoretic models for joint routing, network coding, and scheduling problem. First routing and network coding are modeled by using a new approach based on compressed topology matrix that takes into account the inherent multicast gain of the network. The scheduling is optimized by a new approach called *network graph soft coloring*. Soft graph coloring is designed by switching between different *components* of a wireless network graph, which we refer to as graph fractals, with appropriate usage rates. The network components, represented by graph fractals, are a new paradigm in network graph partitioning that enables modeling of the network optimization problem by using the matrix game framework. In the proposed game which is a *nonlinear cubic* game, the strategy sets of the players are links, path, and network components. The outputs of this game model are mixed strategy vectors of the second and the third players at equilibrium. Strategy vector of the second player specifies optimum multi-path routing and network coding solution while mixed strategy vector of the third players indicates optimum switching rate among different network components or membership probabilities for optimal soft scheduling approach. Optimum throughput is the value of the proposed *nonlinear cubic* game at equilibrium. The proposed *nonlinear cubic* game is solved by extending fictitious playing method. Numerical and simulation results prove the superior performance of the proposed techniques compared to the conventional schemes using hard graph coloring.

## I. INTRODUCTION

Optimization of routing for unicast networks where each source is sending its data to just one sinks, is straightforward and many algorithms are already available on the topic [1-5]. Multicasting, which is sending data from the source to a group of destinations, provides multicast gain resulting into decreased overall required network resources [6-9]. In addition to this inherent gain multicasting provides a chance for using network coding gain. Although network coding can be also used in unicasting with multiple sessions, its main advantage is for multicasting. Network coding with additional process at relays and even source nodes increases the overall throughput. This technique, originally proposed for satellite communications, is extended for applications in wired and wireless communications [10-14]. Although plain routing and network coding usually are not used together, a combination of these techniques can improve the performance [15-17]. When the problem of joint routing and network coding is extended to wireless ad-hoc network, interference between links must be controlled through scheduling, because of omnidirectional transmissions. Scheduling in wireless communications is modeled as a coloring problem in graph theory where two adjacent areas (sub graphs) cannot be painted with the same color [18-23]. Therefore the problem of routing and network coding should be solved simultaneously with scheduling [24-27].

In this paper a *nonlinear cubic* game is proposed to jointly *optimize* routing, network coding, multicast gain and scheduling. This new approach optimizes scheduling by introducing concept of soft coloring where any link can be painted with more than one distinct color. The proposed technique although needs to solve a complex non-linear game, presents *the optimal solution* for the problem. On the other hand, dominancy theory helps us to reduce volume of the cubic payoff matrix.

To solve this game model we propose an extended fictitious play (FP) technique. In FP players can update their belief on other player's strategies based on history of their decisions. This technique which is used to solve matrix games was proposed by Brown [28] and its convergence proved in [29]. FP has originally been proposed for ordinary linear games and in this paper we extend its application for the proposed *non-linear cubic* game model.

### A. Review of Related Works

Reference [30] introduces the application of matrix games to find capacity of wired networks with network switching and [17] extends its application to optimize joint routing and network coding for wired networks. In [30] and [17], presentation of graph is modified to path matrix instead of conventional definition by set of nodes and links. Reference [31] uses this problem formulation for joint optimization of routing, and network coding in wireless networks and solves it with a heuristic algorithm. For a large wireless network, path matrix has excessively large dimensions and application of matrix games helps us to simplify and solve joint optimization problem.

References [32]-[36] are of particular relevance for our work. For this reason in the sequel we will explicitly point out the difference between our work and the results presented in these papers.

Reference [32] discusses network coding games with *unicast* flows. To implement network coding, users need to coordinate and cooperate with respect to their strategies in terms of duplicating and transmitting side information across specific parts of the network. In unicast applications where users have no inherent interest in providing (or concealing) their information to (or from) any destinations except for their unique one, this assumption becomes critical in the face of users' autonomy. A trivial approach to unicast network coding would be for intermediate nodes to perform network coding whenever possible. Via a simple generalization of the two user butterfly network, [32] shows that in general this approach is not desirable; it not only fails to achieve capacity, but it may also induce a network coding game with undesirable equilibria. [32] constructs a network coding scheme for an example that not only achieves capacity (implying a Pareto optimal solution), but does so in a dominant strategy sense. In other words, the proposed network coding scheme ensures that the desirable capacity achieving solution emerges as a dominant strategy equilibrium point of the game.

Reference [33] also uses a game theoretic framework for studying a restricted form of network coding in a general wireless network. The network is fixed and known, and the system performance is measured as the number of wireless transmissions required to meet $n$ unicast demands. Game theory is here employed as a tool for improving distributed network coding solutions. A framework that allows each unicast session to independently adjust his routing decision in response to local information is proposed. The unicast sessions are modeled as self-interested decision makers in a noncooperative game.

This approach involves designing both local cost functions and decision rules for the unicast sessions so that the resulting collective behavior achieves a desirable system performance in a shared network environment. The performance of the resulting distributed algorithms is compared to the best performance that could be found and implemented using a centralized controller.

The focus is on the performance of stable solutions defined as a form of Nash equilibrium. Results include bounds on the best- and worst-case stable solutions as compared to the optimal centralized solution. The result show that bounds on the best- and worst-case stable performance cannot be improved using cost functions that are independent of the network structure. Results in learning in games prove that the best-case stable solution can be learned by self-interested players with probability approaching 1.

In [34] the contents of network flows are specified through network coding (or plain routing) in network layer and the throughput rates are jointly optimized in medium access control layer over fixed set of conflict-free transmission schedules (or optimized over transmission probabilities in random access). If the network model incorporates bursty sources and allows packet queues to empty, the objective is to specify the stability region as the set of maximum throughput rates that can be sustained with finite packet delay. Dynamic queue management strategies are used to expand the stability region toward the maximum throughput region. Because of the complexity introduced by the presence of multiple sources, the requirement of stable operation, and the wireless network properties, [34] restricts its attention to a simple *tandem* network topology with at most two-node connectivity such that any packet transmission reaches only its left and right neighbors, respectively which oversimplifies the routing problem.

In [35] a method is proposed to identify structural properties of multicast network configurations, by decomposing networks into regions through which the same information flows. This decomposition demonstrates that very different networks are equivalent from a coding point of view, and offers a means to identify such equivalence classes. It also allows us to divide the network coding problem into two almost independent tasks: one of graph theory and the other of classical channel coding theory. This approach to network coding enables us to derive the smallest code alphabet size sufficient to code any network configuration with two sources as a function of the number of receivers in the network. [35] proposes deterministic algorithms to specify the coding operations at network nodes without the knowledge of the overall network topology. Such decentralized designs facilitate the construction of codes that can easily accommodate future changes in the network, e.g., addition of receivers and loss of links. In our work network topology decomposition is used for scheduling purposes.

The rest of the paper is organized as follows. In Section II, the system model for wireless ad-hoc networks is presented. The matrix game framework combined with soft graph coloring for optimization of link scheduling is presented in Section III. In Section IV, the proposed matrix game model is extended as a *nonlinear cubic* game to jointly optimize routing, network coding, multicast gain and scheduling along with extended FP algorithm to solve the proposed *nonlinear cubic* game is presented. In Section V, numerical results are presented and finally paper is concluded in Section VI.

## II. PROBLEM DEFINITION AND SYSTEM MODEL

Assume a multi-source wireless *ad-hoc* network including $N$ nodes. This network is defined as $G(V, E, \zeta, s_1, s_2, \ldots, s_M)$ where $V$ is set of nodes with $N$ elements, $E$ is set of $L$ wireless links, $\zeta$ is set of $M$ sources and $s_i$ is set of sinks corresponding to the $i$th source. If source is sending unicast data, size of its sink set is one. Wireless propagation for this network assumes the following:

1. Omni-directional transmission.
2. Presence of interference due to simultaneous transmission.
3. TDMA as multiple access scheme for different hops without inter time slot interference.

### A. Conflict Free Operation

Assume $S_{ij}$ as the power, in dB, required at node $j$, for transmitting node $i$ to reach the receiving node $j$ at distance $d_{ij}$ with $S_{ij} \propto S_i d_{ij}^{-\alpha}$ where $\alpha$ is attenuation factor and $S_i$ is transmission power of $i$th node.

By, definition of the conflict free scheduling [37,38], any node $k \neq i, j$, receiving the signal from node $m$, will be interfered by link $l_{ij}$ if and only if $S_{mk} \leq S_{ik} + \beta$, where $\beta$, in dB, is acceptable interference margin between two links. In other word, link $l_{ij}$ is adjacent to $l_{mk}$ for any $m$ and any $k \neq i, j$ if

$$S_{mk} \leq S_{ik} + \beta. \tag{1}$$

Alternatively, the two links are adjacent if

$$S_{ij} \leq S_{mj} + \beta. \tag{2}$$

Whenever $l_{ij}$ and $l_{mk}$ are physically adjacent i.e. they have a common node or either (1) or (2) hold, they cannot be painted by the same color. Using (1) and (2), link adjacency matrix which is used to design network graph coloring algorithm is defined. This type of conflict free transmission needs exact knowledge of the distance between any two nodes in the network.

A simpler definition for conflict free transmission counts distance between links in terms of number of hops and usually if distance between two links is more than one hop they are assumed to be not-adjacent. Two links are called $k$ hops-adjacent if there are connected through $k-1$ links and when two links are physically adjacent they are called one hop-adjacent.

*B. Paths Identification*

The next step is to identify the set of all possible paths from each source to its corresponding set of sinks. Unicast path identification can be done using any algorithm available in the literature. For instance Dijkstra algorithm finds shortest path and it can be easily extended to find $I^{m,n}$ shortest path from $m$th source to its $n$th corresponding sink.

Assume $P_i^{m,n}$ as $i$th non-cyclic directed path from the $m$th source to its $n$th corresponding sink. Set of all multicast paths from source m to the set of sinks $\mathbf{s}_m$ are calculated as follows,

$$P^m_{i_1,i_2,\cdots,i_{|s_m|-1},i_{|s_m|}} = \bigcup_{n=1}^{|s_m|} P^{m,n}_{i_n}, \qquad (3)$$

where $\bigcup$ is union and $|.|$ is cardinality operator. We also define network coded paths as sets of links to carry minimum required systematic and parity data from sources to sinks whereas received packets at sinks are decodable. Assume simple butterfly network Fig. 1 where first node is source and last two nodes are multicast sinks. We can define 9 multicast paths as follows,

$$P_1 = P_{1,1}^I = \{l_{12}, l_{25}, l_{13}, l_{36}\}, P_2 = P_{1,2}^I = \{l_{12}, l_{25}, l_{24}, l_{46}\}, P_3 = P_{1,3}^I = \{l_{12}, l_{25}, l_{13}, l_{34}, l_{46}\},$$

$$P_4 = P_{2,1}^I = \{l_{12}, l_{24}, l_{45}, l_{13}, l_{36}\}, P_5 = P_{2,2}^I = \{l_{12}, l_{24}, l_{45}, l_{46}\}, P_6 = P_{2,3}^I = P_{3,2}^I = \{l_{12}, l_{24}, l_{45}, l_{13}, l_{34}, l_{46}\},$$

$$P_7 = P_{3,1}^I = \{l_{13}, l_{34}, l_{45}, l_{36}\}, P_8 = P_{3,3}^I = \{l_{13}, l_{34}, l_{45}, l_{46}\}, P_9 = P_{NC} = \{l_{12}, l_{13}, l_{24}, l_{25}, l_{34}, l_{36}, l_{45}, l_{46}\}$$

where last path is network coded path and contains all links of butterfly network.

Next step is compression of the selected paths due to omnidirectional nature of the propagation and inherent multicast gain. If one multicast path consists of two or more links carrying the same information and originating from the same source, they can be substituted with one equivalent link dependent on the definition of the adjacency. If link adjacency is defined based on interference margin, those links are replaced with the one with higher power. If *k*-hops adjacency definition is used, those links are substituted with one new link. We refer to this process as topology compression that reduces the network interference graph and simplifies the scheduling process. For instance for topology in Fig. 1 and using *k*-hops adjacency definition, $P_1$, $P_2$,…, $P_9$ are compressed as:

$$P_1 = \{l_{1(2,3)}, l_{25}, l_{36}\}, P_2 = \{l_{12}, l_{2(4,5)}, l_{46}\}, P_3 = \{l_{1(2,3)}, l_{25}, l_{34}, l_{46}\}, P_4 = \{l_{1(2,3)}, l_{24}, l_{45}, l_{36}\},$$

$$P_5 = \{l_{12}, l_{24}, l_{4(5,6)}\}, P_6 = \{l_{1(2,3)}, l_{24}, l_{4(5,6)}, l_{34}\}, P_7 = \{l_{13}, l_{3(4,6)}, l_{45}\}, P_8 = \{l_{13}, l_{34}, l_{4(5,6)}\},$$

$$P_9 = \{l_{12}, l_{13}, l_{2(4,5)}, l_{3(4,6)}, l_{4(5,6)}\},$$

where for instance new link $l_{1(2,3)}$ means link originating from node 1 and ending at both nodes 3 and 5 realized with only one transmission. With this notation, the problem of optimum multicast routing is finding optimum usage rates for the identified paths.

*C. Conventional Scheduling: Motivating Example*

In general conflict free scheduling is based on conventional network graph coloring techniques. When conventional graph coloring is used, any link is painted by just one color and no two adjacent links are painted with the same color. When all links have the same capacity and they are used with the same rate, any minimal conventional coloring is optimum solution for links scheduling. Following simple example shows inefficiency of coloring when links must be activated with different usage rates. Assume that we have 3 links $l_1$, $l_2$, $l_3$ with usage rates $r_1=3$, $r_2=1$, $r_3=2$. This means that within a given time frame, referred to as clique cycle to be minimized, link $l_1$ is used during 3 time slots ($r_1=3$), link $l_2$ during one slot only ($r_2=1$) and link $l_3$ during 2 slots ($r_3=2$). In addition, due to a given link adjacency, link 1 can be activated with other two simultaneously and two others cannot be activated together. The minimal coloring schemes for these 3 links with their assigned rates are as follows

i) $T_1=\{l_1, l_2\}$ *and* $T_2=\{l_3\}$ required number of time slots (clique cycle) is 5.

ii) $T_1=\{l_1, l_3\}$ *and* $T_2=\{l_2\}$ required number of time slots is 4.

Obviously none of these coloring schemes is optimum and the optimum scheduling for these links is,

iii) $T_1=\{l_1, l_2\}$ *during the first time slot and* $T_2=\{l_1, l_3\}$ *during the next two slots;* required number of time slots is 3.

Therefore in optimum solution $T_1$ and $T_2$ have non-empty intersection and we call non-overlapped partial topologies as components of the topology represented by *graph fractals*. A topology or network component set is a complete set of links that can be activated at the same time. Any single link is also a component and we call them as first generation network components and $\tau$th generation of network components refers to the set of all components with $\tau$ members. Network components of each generation are parents of the next generation. For our 3-links example, we have 5 components as follows, $T_1=\{l_1\}$,

$T_2=\{l_2\}$, $T_3=\{l_3\}$, $T_4=\{l_1, l_2\}$ and $T_5=\{l_1, l_3\}$, where first 3 components are parents of the last two ones. Consequently in general case to optimize the scheduling appropriate network components and their optimal usage rates must be found.

### III. MATRIX GAME MODELING FOR OPTIMUM SCHEDULING

*A. Formulation of the Game*

Given link activation rate *r* and link capacity *c* vectors with *I* elements and network component set $\zeta$ with *J* elements, we define a payoff matrix *H* as follows,

$$h_{ij} = \begin{cases} \dfrac{c_i}{r_i}, & \text{if } l_i \in \zeta_j \\ 0, & \text{otherwise} \end{cases} \qquad (4)$$

where $r_i$ and $c_i$ are activation rate and capacity respectively for *i*th link and $\zeta_j$ is the *j*th network component. Assume *H* as payoff matrix of the min-max zero sum game between two players where their strategy sets are links and network components sets respectively. In the sequel we prove that the mixed strategy vector of the second player gives optimum network component rate.

*Theorem 1-* mixed equilibrium of the zero sum game defined by payoff matrix (4) gives optimum scheduling for usage rate vector *r*.

Proof: Assume *x* and *y* as mixed strategies of the players at equilibrium. Since *y* is normalized to 1, activation of network components in average needs one time slot. Parameter $\tilde{r}_i$ as activation rate of *i*th link supported by components of vector *y* is computed as follows,

$$\frac{\tilde{r}_i}{r_i} =_i [H]y \qquad (5)$$

where $_i[H]$ is $i$th row of the $H$. Therefore for a given component rate vector $y$, the link with minimum supported rate, which is bottleneck of the game is,

$$i_{min\,imum} = \arg\min{}_i \left(_i[H]y\right), \qquad (6)$$

Consequently optimum component rate vector must maximize $\min{}_i \left(_i[H]y\right)$ as

$$y = \arg\max{}_y \min{}_i \left(_i[H]y\right). \qquad (7)$$

And theoretically (7) is equivalent to

$$x = \arg\min{}_x \left(x^T H y\right), \qquad (8)$$

$$y = \arg\max{}_y \left(x^T H y\right), \qquad (9)$$

where (8) and (9) define equilibrium for game defined by payoff matrix $H$ and value

$$\text{Throughput} = \text{Value} = \max{}_y \min{}_x \left(x^T H y\right). \qquad (10)$$

Formulation of the problem as a game gives us the chance to simplify the model using some special properties of the games, like dominancy theory. For instance in general case while solving this game, because of dominancy, components tagged as parent can be ignored because apparently any parent is dominated by its child. Therefore we just need to consider last generation generated (born) from any link. In addition, calculation of mixed strategy for linear games is easy and can be performed by fictitious playing method (FP).

*B. FP Algorithm to Solve the Proposed Game Model*

FP property of the games is a feature of stationary games where players can update their belief on the other player's strategies based on history of their decisions. This technique which is used to solve matrix games, was proposed by Brown [12] and its convergence for different conditions was proved in [13-18]. FP algorithm for a min-max zero-sum game has following steps,

*Algorithm 1: FP algorithm for min-max game*

Step 1. Initialization of $\hat{x}_k$ which is estimated $x$ at $k$th iteration as,

$$\hat{x}_0 = {}_i[H]^T, \qquad (11)$$

where ${}_i[H]$ is and arbitrary row of $H$ ($\hat{x}$ actually will be $Hy$). Then set iteration number $k=1$.

Step 2. Finding best strategy for the first player at $k$th iteration as,

$$i_k = \arg\min_m \hat{x}_{k-1,m} \qquad (12)$$

where $\hat{x}_{k,m}$ is $m$th element of the $\hat{x}_k$.

Step 3. Updating $\hat{y}_k$ as, $\hat{y}_k = \hat{y}_{k-1} + [H]_{i_k}$.

Step 4. Finding best strategy for the second player at $k$th iteration as,

$$j_k = \arg\max_n \hat{y}_{k,n} \qquad (13)$$

where $\hat{y}_{k,n}$ is $n$th element of the $\hat{y}_k$.

Step 5. If the algorithm has not $\delta$ converged to equilibrium where $\delta$ is an small threshold set $k \leftarrow k+1$ then update $\hat{x}_k$ as,

$$\hat{x}_k = \hat{x}_{k-1} + {}_{j_k}[H], \qquad (14)$$

and then return to Step 2. Algorithm is $\delta$ converged if,

$$\hat{y}_{k,j_k} - \hat{x}_{k,i_k} \leq \delta, \tag{15}$$

When the algorithm approaches equilibrium mixed strategies for both players are calculated as probabilities of the strategies taken at all iterations i.e. averaging over $i_k$ s and $j_k$ s.

*C. Numerical Results for Optimum Scheduling*

We present numerical results for network coded butterfly network shown in Fig. 1. For this network path $P_9$ in Section II.B, has 6 links $\boldsymbol{L} = [l_{12}, l_{13}, l_{1(2,3)}, l_{2(4,5)}, l_{3(4,6)}, l_{4(5,6)}]^T$ involved in the path. When these links have the same capacity, conventional and optimum scheduling are the same because they also have the same link usage rate. Therefore we present the result for the case when links have different capacities. Assume $\boldsymbol{C}$ as vector of link capacity rates as,

$$\boldsymbol{C} = [1.2844, 0.7916, 0.7916, 1.1035, 1.4644, 0.6325]^T, \tag{16}$$

where capacity of $l_{1(2,3)}$ is minimum capacity of $l_{12}$ and $l_{13}$. Adjacency matrix for these 6 links is,

$$\boldsymbol{A} = \begin{bmatrix} 1 & 1 & 1 & 1 & 0 & 0 \\ 1 & 1 & 1 & 1 & 0 & 0 \\ 1 & 1 & 1 & 1 & 0 & 1 \\ 0 & 1 & 1 & 1 & 1 & 1 \\ 0 & 0 & 0 & 1 & 1 & 1 \\ 1 & 0 & 1 & 1 & 1 & 1 \end{bmatrix} \tag{17}$$

based on adjacency matrix $\boldsymbol{A}$, 11 network components are defined where first 6 are single link parents and dominated by the last 5. Dominating components are as follows,

$C_1=\{l_{12}, l_{3(4,6)}\}$, $C_2=\{l_{12}, l_{4(5,6)}\}$, $C_3=\{l_{13}, l_{2(4,5)}\}$, $C_4=\{l_{13}, l_{4(5,6)}\}$, $C_5=\{l_{1(2,3)}, l_{4(5,6)}\}$.

After solving the defined game using (8)-(9) and running FP with 2000 iterations, $x$ and $y$ at equilibrium have values as follows,

$$x = [0.2245, 0.3671, 0.3671, 0.0121, 0.0099, 0.0193]^T, \qquad (18)$$

$$y = [0.2256, 0.0104, 0.2942, 0.0875, 0.3823]^T. \qquad (19)$$

Fig. 2 presents transient behaviour of the FP where black and blue curves are payoff received by the first and second players respectively. We can see the payoff received by the second player faces a sharp peak at initial iteration but this peak is disappeared after a few iteration and both payoffs are converged to the same value. On the other hand, as we are expecting payoff received by the second player is always more than one received by the first player. At equilibrium both payoffs approach value of the game which is throughput equal to 0.61 packet/time slot. At the same time for a conventional scheduling with following 3 partial topologies, $T_1=\{l_{12}, l_{3(4,6)}\}$, $T_2=\{l_{13}, l_{2(4,5)}\}$, $T_3=\{l_{4(5,6)}\}$, the achieved throughput is 0.55 packet/time slot. The throughput gain of the proposed optimum scheduling compared to the conventional one is about 11 percent.

## IV. MATRIX GAME MODELING FOR JOINT OPTIMUM ROUTING, NETWORK CODING AND SCHEDULING

### A. Formulation of the Game

Assume $z$ as vector of path usage rates in the problem of multipath routing. In this case, we modify the proposed game model in Section III by adding a third player with mixed strategy vector $z$. Clearly, link usage rates are linear combinations of path usage rates and therefore according to (4), the proposed 3 player game will be defined by modifying parameter $r_i$ in (4) by a linear combination of elements of vector $z$. Consequently, the proposed game in general form becomes *nonlinear cubic* game where strategy sets of its players are links, paths, and network components respectively and the actual payoff and cost received by the first and third players respectively is a nonlinear function of mixed strategy chosen by the second player. According to (4), for a given $z$, if the first and the second players choose their $i$th and $j$th pure strategy, their payoff and cost will be $1/\sum_{k=1}^{K} a_{ijk} z_k$, where $a_{ijk}$ as elements of 3 dimensional (cubic) matrix is defined as follows,

$$a_{ijq} = \begin{cases} 1, & \text{if } l_i \in P_j, l_i \in \kappa_q \text{ and } P_j \text{ is a compressed multicast path,} \\ \dfrac{1}{C_c}, & \text{if } l_i \in P_j, l_i \in \kappa_q \text{ and } P_j \text{ is a compressed network coded path,} \\ \infty, & \text{if } l_i \notin P_j, \\ 0, & \text{Otherwise.} \end{cases} \qquad (20)$$

and $C_c$ is the number of XOR combined links at the node with network coding. For our general problem of joint optimal modeling of the routing, scheduling, and network coding using game theory, set of payoff matrices introduced in the Section III are replaced by the same number of matrices (this time 3 dimensional) with elements $a_{ijk}$. The mixed strategies and value of this game are calculated as follows,

$$\boldsymbol{x} = \arg\min_{x} \sum_{i=1}^{L}\sum_{j=1}^{J} \dfrac{x_i y_j}{\sum_{k=1}^{K} a_{ijk} z_k}, \qquad (21)$$

$$\boldsymbol{y} = \arg\max_{y} \sum_{i=1}^{L}\sum_{j=1}^{J} \dfrac{x_i y_j}{\sum_{k=1}^{K} a_{ijk} z_k}, \qquad (22)$$

$$\boldsymbol{z} = \arg\max_{z} \sum_{i=1}^{L}\sum_{j=1}^{J} \dfrac{x_i y_j}{\sum_{k=1}^{K} a_{ijk} z_k}, \qquad (23)$$

$$\text{Throughput=Value}= \min_x \max_y \max_z \sum_{i=1}^{L}\sum_{j=1}^{J} \frac{x_i y_j}{\sum_{j=1}^{J} a_{ijk} z_k}. \tag{24}$$

Since the value of this game is a linear function of mixed strategy vector of the first and second player, dominancy theory is still applicable for their strategies but this property of linear games is not valid for the third player's strategies.

*Theorem 2*- Dominancy theory for the first two players, in the proposed *nonlinear cubic* game: Strategy $i_1$ of the first player is dominated by its strategy $i_2$ if for any $j$ and $k$, $a_{i_1 jk} \leq a_{i_2 jk}$ and in the same way, strategy $j_1$ of the second player is dominated by its strategy $j_2$ if for any $i$ and $k$, $a_{ij_1 k} \geq a_{ij_2 k}$.

Proof- We only need to prove the first part of the theorem and the second part is proved in the same way. Assume $x$, $y$, and $z$ as strategy vectors at equilibrium. If $a_{i_1 jk} \leq a_{i_2 jk}$ for any $j$ and $k$, therefore $\sum_{k=1}^{K} a_{i_1 jk} z_k \leq \sum_{k=1}^{K} a_{i_2 jk} z_k$ and consequently,

$$val = \sum_{i=1}^{L}\sum_{j=1}^{J} \frac{x_i y_j}{\sum_{k=1}^{K} a_{ijk} z_k} = \sum_{\substack{i=1, j=1 \\ i \neq i_1, \\ i \neq i_2}}^{L,J} \frac{x_i y_j}{\sum_{k=1}^{K} a_{ijk} z_k} + \sum_{j=1}^{J} \frac{x_{i_1} y_j}{\sum_{k=1}^{K} a_{i_1 jk} z_k} + \sum_{j=1}^{J} \frac{x_{i_2} y_j}{\sum_{k=1}^{K} a_{i_2 jk} z_k}, \tag{25}$$

and,

$$val \geq \sum_{\substack{i=1, j=1 \\ i \neq i_1, \\ i \neq i_2}}^{L,J} \frac{x_i y_j}{\sum_{k=1}^{K} a_{ijk} z_k} + \sum_{j=1}^{J} \frac{(x_{i_1} + x_{i_2}) y_j}{\sum_{k=1}^{K} a_{i_2 jk} z_k} \tag{26}$$

Therefore the payoff received by the first player is less than or equal to the value of the game if the player chooses strategy $i_2$ instead of $i_1$ and consequently $i_1$ strategy is dominated by $i_2$. Inequality is possible iff $x_{i_1} = 0$.

### B. Extended Fictitious Playing for Nonlinear Cubic Game (20-22)

To solve (21)-(23), we first prove that against any mixed strategies of the second and third players, the best option of the first player is a pure strategy.

*Theorem 3-* For given mixed strategy vectors $y$ and $z$, first player receives minimum payoff by a pure strategy and for given mixed strategy vectors $x$ and $z$, second player receives maximum payoff by a pure strategy.

*Proof-* Assume, $i_{min} = \arg\min_i \sum_{j=1}^{J} \dfrac{y_j}{\sum_{k=1}^{K} a_{ijk} z_k}$. Consequently we have

$$\sum_{i=1}^{L}\sum_{j=1}^{J} \dfrac{x_i y_j}{\sum_{k=1}^{K} a_{ijk} z_k} \geq \sum_{i=1}^{L}\sum_{j=1}^{J} \dfrac{x_i y_j}{\sum_{k=1}^{K} a_{i_{max} jk} z_k} = \sum_{i=1}^{L} x_i \sum_{j=1}^{J} \dfrac{y_j}{\sum_{k=1}^{K} a_{i_{max} jk} z_k}. \tag{27}$$

Considering $\sum_{i=1}^{I} x_i = 1$, right hand side of (27) is simplified as,

$$\sum_{i=1}^{L}\sum_{j=1}^{J} \dfrac{x_i y_j}{\sum_{k=1}^{K} a_{ijk} z_k} \geq \sum_{j=1}^{J} \dfrac{y_j}{\sum_{k=1}^{K} a_{i_{max} jk} z_k}, . \tag{28}$$

i.e. payoff received by the first player when it chooses $i_{min}$ th strategy is minimum and proof is complete.

The second part of the theorem for the second player is proved in the same way but there is no such a property for the third player and consequently conventional FP algorithm can not be used for the third player.

*Algorithm 2: Extending Fictitious Playing Algorithm for nonlinear cubic game (21-23).*

Step 1. Initialization of $y$, $z$ as follows,

$q = 0$, $y_j^{(q)} = 0$, for $\forall j \in [2, J]$ and $y_1^{(q)} = 1$, and $z_k^{(q)} = 0$, for $\forall k \in [2, K]$ and $z_1^{(q)} = 1$.

Step 2. Setting iteration number $q \leftarrow q+1$ and calculation of the payoff received by different strategies of the first player as,

$$V_{x,i}^{(q)} = \sum_{j=1}^{J} \frac{y_j^{(q-1)}}{\sum_{k=1}^{K} a_{ijk} z_k^{(q-1)}}, \tag{29}$$

Step 3. Choosing best strategy for the first player against $y^{(q)}$ and $z^{(q)}$ as,

$$i_{min}^{(q)} = \arg \min_i V_{x,i}^{(q)}, \tag{30}$$

Step 4. Updating $x^{(q)}$ as $x_{i_{min}^{(q)}}^{(q)} \leftarrow x_{i_{min}^{(q)}}^{(q-1)} + 1$ and $x_i^{(q)} \leftarrow x_i^{(q-1)}$ for $\forall i \neq i_{min}^{(q)}$.

Step 5. Calculation of the payoff received by different strategies of the second player as,

$$V_{y,j}^{(q)} = \sum_{i=1}^{I} \frac{x_i^{(q)}}{\sum_{k=1}^{K} a_{ijk} z_k^{(q-1)}}, \tag{31}$$

Step 6. Choosing best strategy for the second player against $x^{(q)}$ and $z^{(q)}$ as,

$$j_{max}^{(q)} = \arg \max_j V_{y,j}^{(q)}, \tag{32}$$

Step 7. Updating $y^{(q)}$ as $y_{j_{max}^{(q)}}^{(q)} \leftarrow y_{j_{max}^{(q)}}^{(q-1)} + 1$ and $y_j^{(q)} \leftarrow y_j^{(q-1)}$ for $\forall j \neq j_{max}^{(q)}$.

Step 8. Updating $z^{(q)}$ using following iterative algorithm which is based on steepest descent algorithm. (Notice, value of the game is a non-linear function of z and consequently best strategy of the third player at iteration q can not be found by just maximization over strategy set of this player.)

8.2. Initialization $z_0^{(q)}$ as $z_0^{(q)} = \dfrac{z^{(q-1)}}{\left|z^{(q-1)}\right|}$ and go to the first iteration $t=1$.

8.2. Updating $z_t$ as

$$\tilde{z}_{t,k}^{(q)} = z_{t-1,k} - \mu \sum_{i=1}^{I}\sum_{j=1}^{J} \dfrac{x_i^{(q)} y_j^{(q)}\left(a_{ijk} - a_{ijK}\right)}{\left(\sum_{k=1}^{K} a_{ijk} z_{t-1,k}^{(q)}\right)^2 \sum_{i=1}^{I} x_i^{(q)} \sum_{j=1}^{J} y_j^{(q)}}, \quad \text{for } k \neq K \tag{33}$$

$$\tilde{z}_{t,k}^{(q)} = z_{t-1,k} + \mu \sum_{k=1}^{K-1}\sum_{i=1}^{I}\sum_{j=1}^{J} \dfrac{x_i^{(q)} y_j^{(q)}\left(a_{ijk} - a_{ijK}\right)}{\left(\sum_{l=1}^{K} a_{ijk} z_{t-1,l}^{(q)}\right)^2 \sum_{i=1}^{I} x_i^{(q)} \sum_{j=1}^{J} y_j^{(q)}}, \quad \text{for } k = K \tag{34}$$

where $\mu$ is step size.

8.3. Assuming $\rho$ as minimum of $\tilde{z}_{t,k}$ versus $k$, $z_{t,k}$ is updated as,

$$z_{t,k} = \begin{cases} \dfrac{\tilde{z}_{t,k}}{\left|\tilde{z}_{t,k}\right|}, & \text{if } \rho \geq 0 \\[2ex] \dfrac{\tilde{z}_{t,k} - \rho}{\left|\tilde{z}_{t,k} - \rho\right|} & \text{if } \rho < 0 \end{cases} \tag{35}$$

8.4. $t \leftarrow t+1$, and return to 8.2 for the next iteration. This sub-algorithm is repeated until have convergence.

Step 9. Updating $z^{(q)}$ as $z^{(q)} = z^{(q-1)} + z_t^{(q)}$.

Step 10. Going to step 2, and repeating steps 2-9 to have convergence.

Step 11. After $Q$ iterations when convergence condition holds calculated mixed strategy vector are normalized to one as $x^{(eq)} = \frac{x^{(Q)}}{|x^{(Q)}|}$, $y^{(eq)} = \frac{y^{(Q)}}{|y^{(Q)}|}$, and $z^{(eq)} = \frac{z^{(Q)}}{|z^{(Q)}|}$. This normalization during iterations before convergence is not necessary and can be performed only once after convergence.

*C. Numerical Results for Nonlinear Cubic Game*

For numerical results, we consider the same scenario as in Section III, i.e. butterfly network shown in Fig. 1, with 8 paths defined in Section II and link capacities and network components set defined in Section III. We solve this joint optimization and routing problem using the proposed FP algorithm for *non-linear cubic* game. For 1-hop adjacency, Paths 4, 5, 6, and 8 are dominated by other 5 paths and therefore we actually need to find rates for 4 remained paths. Fig. 3 shows transient behavior of the payoffs received by 3 players. It can be seen that, for lower iteration index, payoff received by the third player has a sharp peak. Although this peak disappears for higher iteration indices, but it can be avoided by better initialization in the steepest descent algorithm proposed to update mixed strategy of the 3$^{rd}$ player. As we are expecting although payoff received by the first player is always less than ones received by its opponents but all payoff asymptotically converge to the same value. Mixed strategies of 3 players at equilibrium are,

$$x = [0.0025, 0.0055, 0.0015, 0.2928, 0.2245, 0.4733]^T, \quad (36)$$

$$y = [0.1025, 0.0510, 0, 0.0625, 0.7839]^T, \quad (37)$$

$$z = [0.2284, 0.0030, 0.3003, 0.2289, 0.2394]^T. \quad (38)$$

The value of the game which is equivalent to the throughput is 0.76 which show about 25 percent gain compared to single path routing with optimum scheduling and 36 percent gain compared to single path routing with conventional scheduling.

V. SIMULATION RESULTS

Simulation results are presented in two different parts for multiple single-path (unicast) and single source multipath (multicast) routing scenarios respectively. The first part presents simulation results for optimum scheduler presented in Section III and second part presents simulation results for joint routing and scheduling algorithm presented in Section IV. For both parts, network graphs are generated randomly with nodes uniformly distributed over unit square and Monte Carlo simulation technique is used.

*A. Simulation Results for Multiple Single-Path Unicast Sessions*

Game model proposed in Section III with random selection of sources and their corresponding sinks is used. For each source-sink pair shortest path (physically shortest part that requires minimum power) is calculated using Dijkstra algorithm [37]. Attenuation factor $\alpha$ is assumed to be 4 and number of packets to deliver from each source to its corresponding sink is randomly chosen by Poisson distribution.

The proposed optimal algorithm is compared to the conventional scheduling based on network graph hard coloring. The average number of required time slots to carry each packet from source to its corresponding sink is considered as performance criterion. This parameter is reciprocal of the value of the game calculated by (10) (throughput). Results are averaged over 1000 independent runs. At each run, link usage rate is calculated from load on each path and then using (1) and (2) adjacency matrix is defined. Finally using (4) game model is defined and it is solved using FP algorithm presented in Section III. Results are summarized in Figs. 4-5.

Fig. 4, presents average number of required time slots (inverse to the system throughput) versus number of parallel unicast source-sink pairs (different numbers of source-sink sessions) when $N =10$ nodes. We can see from the figure that the proposed scheduling outperforms the conventional scheme. In the case of higher number of source-sink sessions, optimum scheduling offers more gain because when higher number of paths is participating in the scheduling process there are more chances that soft coloring will find the combination that is more efficient than the hard coloring. In other words every link has more potential partners for simultaneous transmission, the number of last generation of components is higher and consequently scheduling gain is also higher. In this case, the proposed optimum scheduling needs between 11 and 25 percent less time slots in average.

Fig. 5, presents the same results for N=20. In this case, we can see that the scheduling gain compared to conventional scheduler is even higher than in Fig. 4.

### B. Simulation Results for Optimization of Multicast Routing and Scheduling

In this part, *nonlinear cubic* game model presented in Section IV, is simulated over randomly generated graphs with uniformly distributed nodes for the cases $N=20$ up to N=200 nodes, and number of sinks between 1 to 6 . Results are averaged over 100 independent runs. In each run, one node is randomly selected as source and some of other nodes are selected as its corresponding sinks. Then using a modified version of Dijkstra algorithm up to 4 shortest paths from source to each sink are identified and then by combining the identified unicast path using (3) set of multicast paths are constructed. In the next step, wireless links involved in the set of multicast paths are identified and then their adjacency matrix is defined using (1) and (2). Finally the non-*linear cubic game* is defined using (20) and solved by using modified FP algorithm presented in Section IV.B and the throughput is calculated using (24). Simulation results are presented for both NC coded and non coded cases combined with both conventional and optimum scheduling and compared with corresponding results for single path case with conventional and optimum scheduling presented in Section III. To decrease the complexity especially for large network sizes where the number of multicast paths exponentially increases we have limited the number of multicast paths to 10 i.e. 10 lowest power path are selected. This simplification does not affect the performance because as we expect at equilibrium just a few the best paths are picked up and coefficient for high power path is always zero.

Fig. 6 presents simulation results for 5 multicast receivers while the interference margin $\beta$=5dB. It can be seen that for all of network sizes, optimized multipath routing outperforms single path case and this improvement when network coding is used is even high. Throughput gain resulting from using network coding in multipath routing is up to 9 percent while for single path routing scheme this gain is negligible. In both NC coded and non coded cases, proposed optimum scheduling improves throughput up to 15.8 percent for a large network. On can see form Fig. 6 that when there are 5 sinks, the overall joint optimization of routing, scheduling, and network coding improves the throughput up to 55 percent.

Fig. 7 presents the same results for $N$=100 nodes versus the number of multicast receivers. As we expect multicast throughput decreases with increasing number of multicast receivers but the total throughput received by all receivers increases. Throughput gain achieved by NC increases with increasing number of multicast sinks (multipath routing).

Fig. 8 presents throughput per normalized power for $N$=100 nodes versus number of multicast receivers to demonstrate that throughput improvement observed in Figs. 7 and 8 does not cost more power consumption. In this case power has been normalized with respect to the dimensions of the environment. In this figure we observe the same behavior for different curves as in Fig. 7. In this case the improvement in throughput per power when there are 6 multicast receivers is 57 percent which is even a bit more than throughput gain in Fig.7.

## VI. CONCLUSIONS

In this paper novel techniques based on matrix games are proposed to solve joint optimal multipath routing and scheduling problem in wireless *ad-hoc* network. In the first proposed technique *soft scheduling* is modeled by matrix games. To define the game model, wireless topology is first partitioned into networks components, represented by *graph fractals,* which are actually overlapped partial topologies. Then by assigning appropriate usage rates to the selected network components optimum scheduling is achieved. The optimal rates for graph fractals are obtained by calculating mix equilibrium of a matrix game between links and network components. Using matrix games, in this case, helps us to simplify the problem by applying dominancy theory and fictitious playing method, which works iteratively, provides quick convergence.

The proposed optimum scheduling is extended to joint optimum multipath routing and scheduling by adding a third player to the scheduling game. Mixed strategy of this new player at equilibrium optimizes rates assigned to each path. The value of the proposed game is a non-linear function of mixed strategy of the third player and we call this game *nonlinear cubic* game. Because of non-linearity, the proposed game model cannot be solved using conventional FP algorithm and therefore we present an efficient modified FP algorithm to solve this problem. Simulation results prove the efficiency of the proposed game models. Using this joint optimization throughput and throughput per power improve up to 55 and 57 percents respectively compared to conventional solutions.

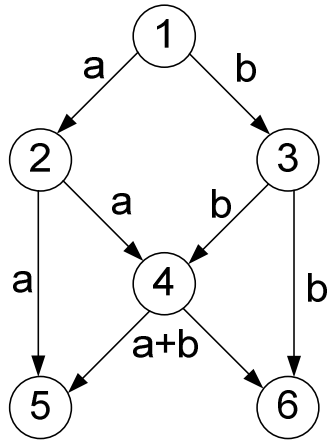

Fig. 1. Typical butterfly network with network coding.

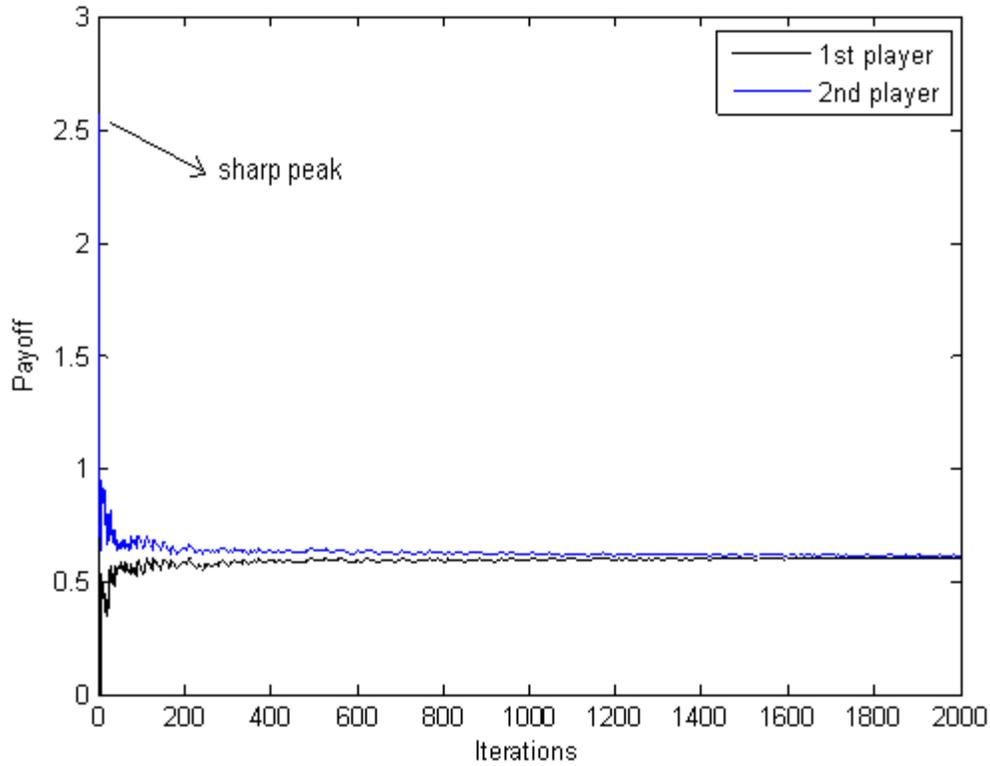

Fig. 2. Payoff received by two players w.r.t. iteration number when scheduling in single path butterfly optimized.

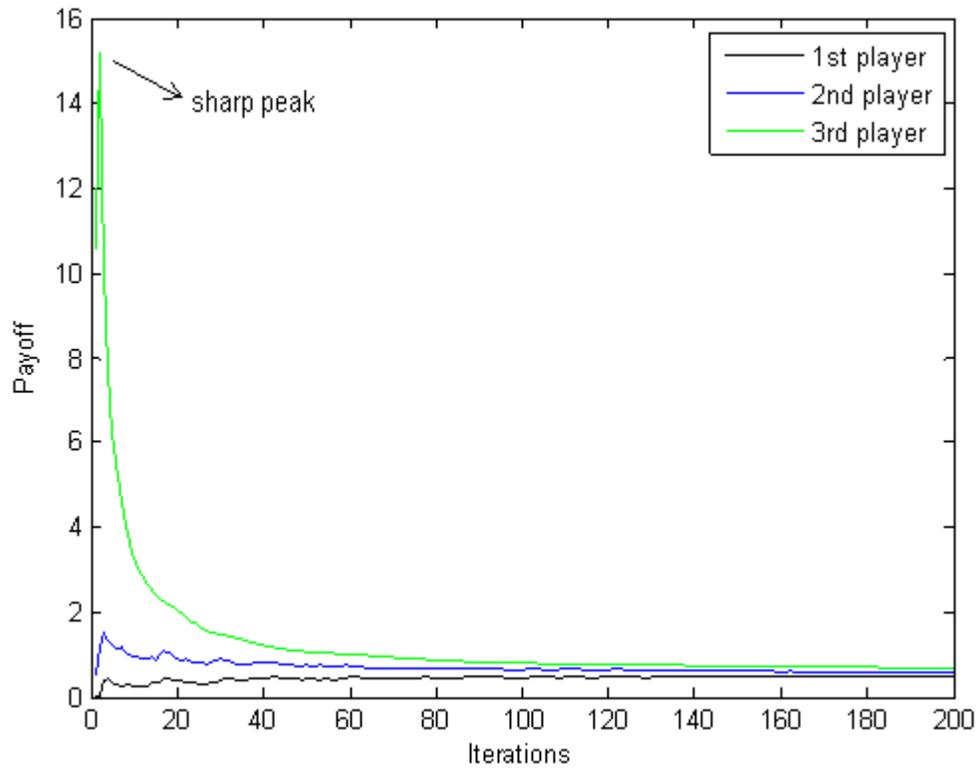

Fig. 3. Payoff received by 3 players w.r.t. iteration number from *non-linear cubic* game when joint routing and scheduling in multipath butterfly network is optimized.

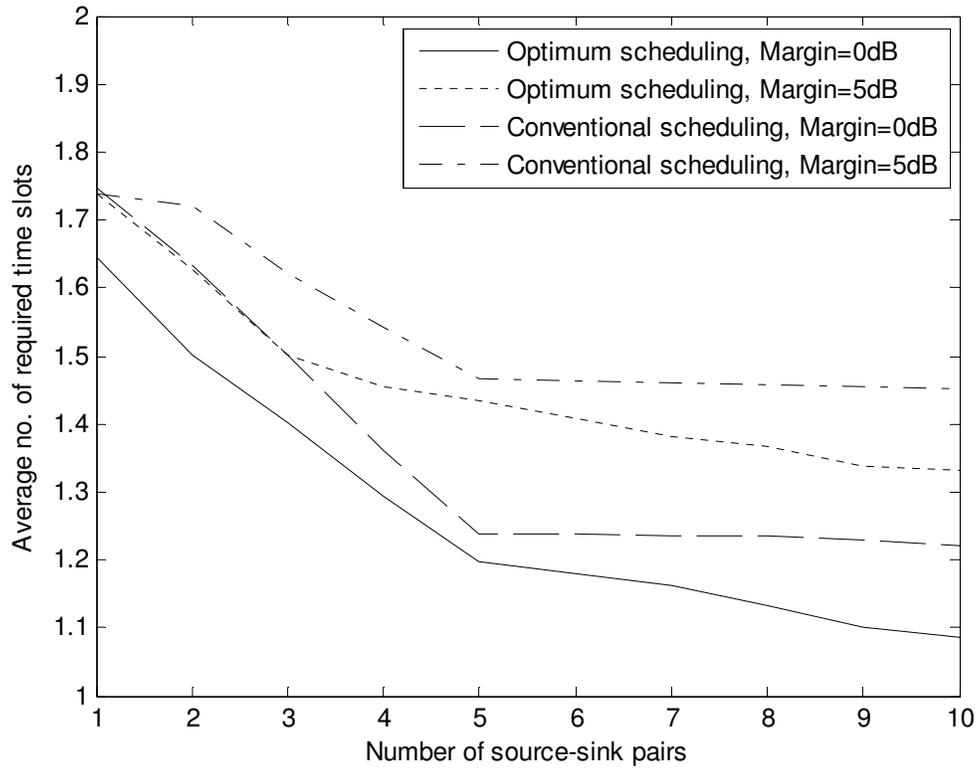

Fig. 4. Average no. of required time slots w.r.t. number of source-sink pairs for *N*=10.

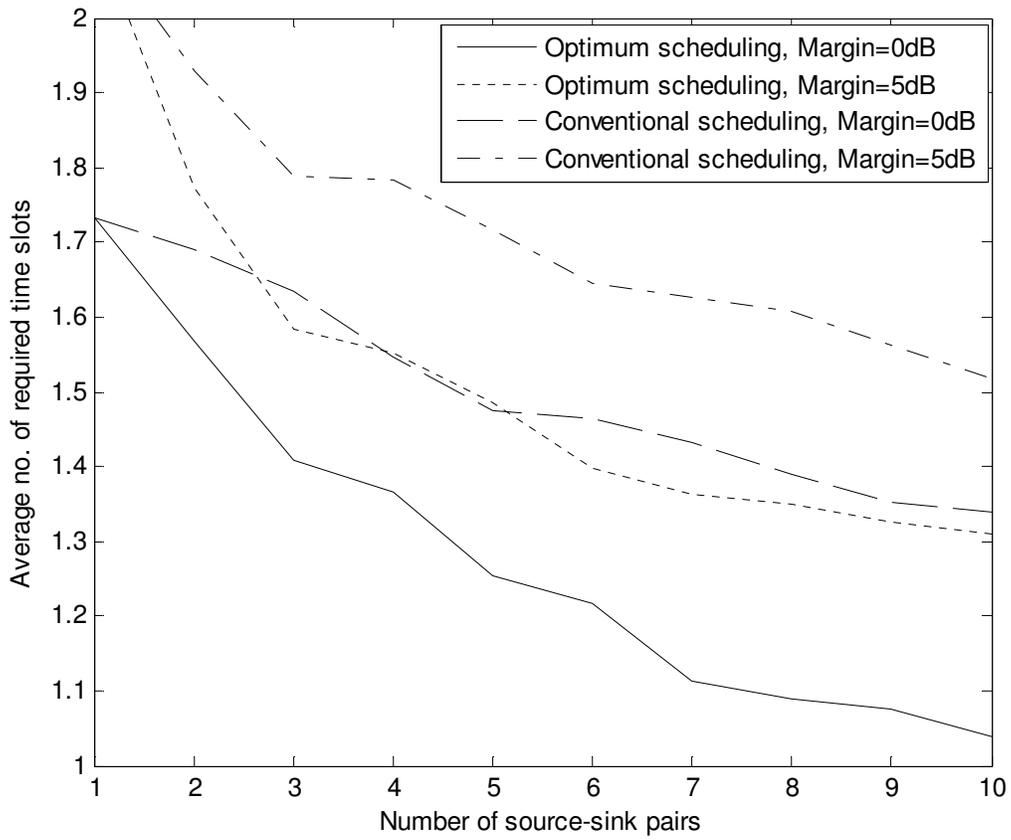

Fig. 5. Average no. of required time slots w.r.t. number of source-sink pairs for *N*=20.

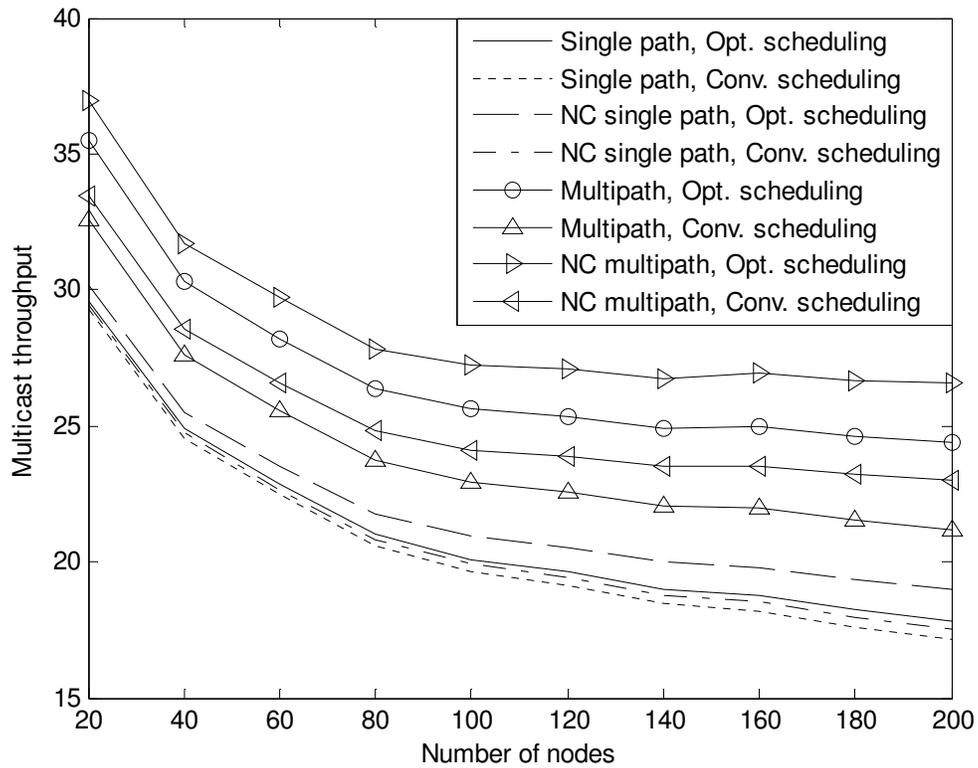

Fig. 6. Multicast throughput of the optimized multipath and single-path routing for both conventional and optimum scheduling with and without network coding w.r.t. number of nodes in the networks while $\beta$=5dB and there are 5 multicast sinks.

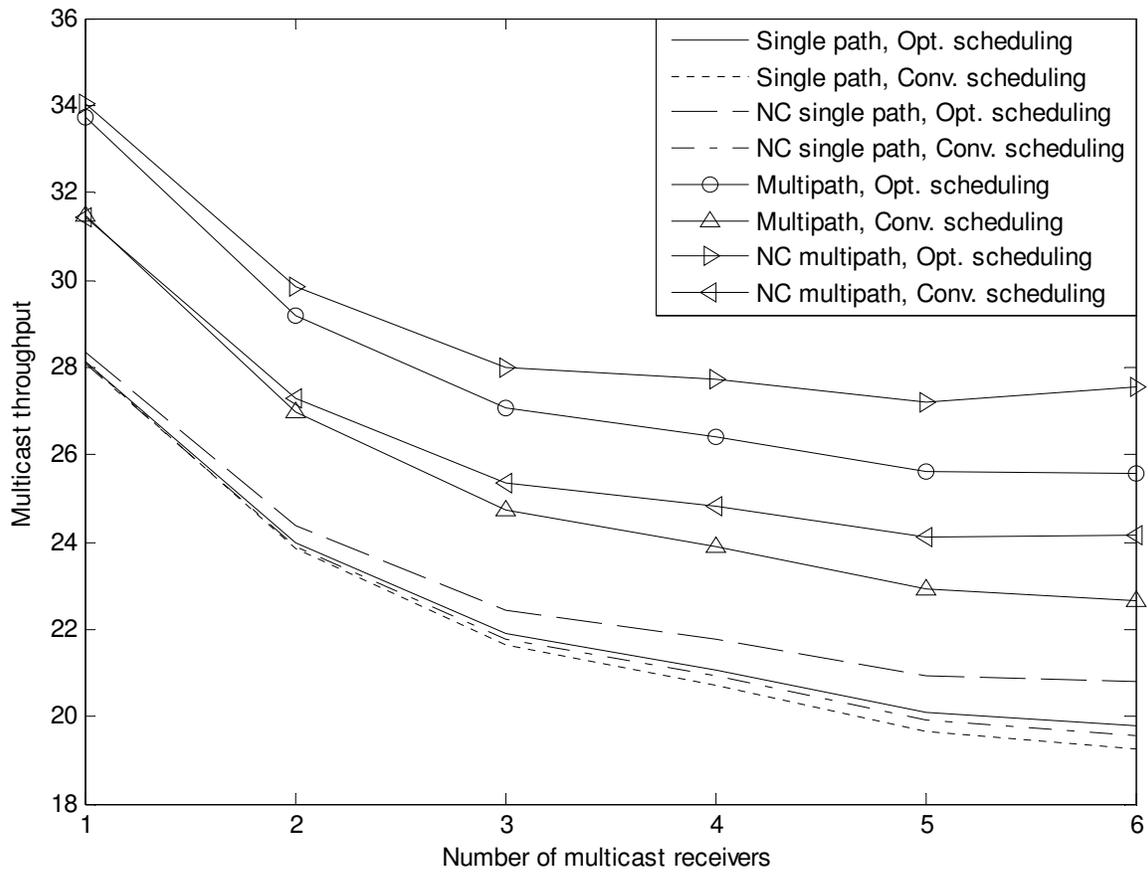

Fig. 7. Multicast throughput of the optimized multipath and single-path routing for both conventional and optimum scheduling with and without network coding w.r.t. number of nodes in the networks while $β$=5dB and there are 100 nodes in the network.

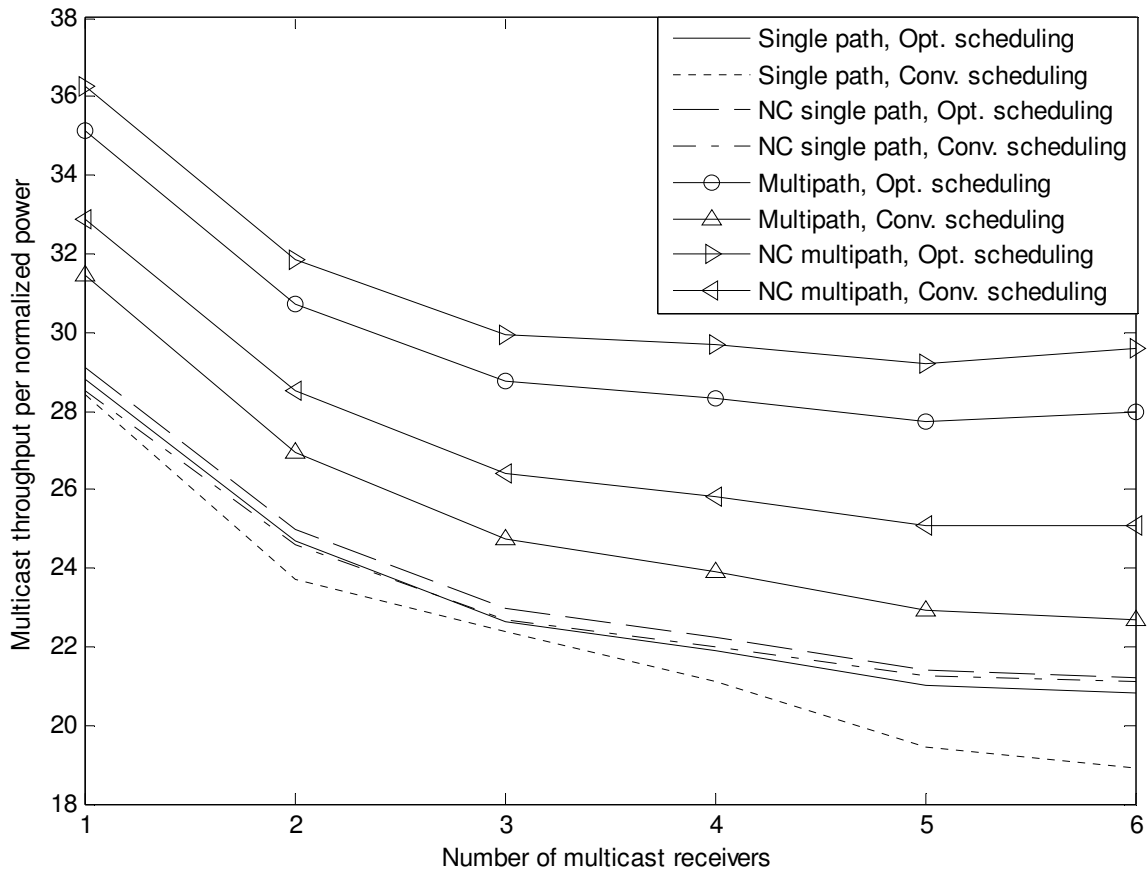

Fig. 8. Multicast throughput per normalized power of the optimized multipath and single-path routing for both conventional and optimum scheduling with and without network coding w.r.t. number of nodes in the networks while $\beta$=5dB and there are 100 nodes in the network.